\documentstyle[twocolumn,epsf,aps]{revtex}
\draft
\begin{document}
\title{Defensive alliances in spatial models of cyclical population
interactions}
\author{Gy\"orgy Szab\'o$^1$ and Tam\'as Cz\'ar\'an$^{2,3}$}

\address{$^1$Research Institute for Technical Physics and Materials Science
POB 49, H-1525 Budapest, Hungary \\
$^2$Theoretical Biology and Ecology Research Group of the Hungarian Academy
of Sciences \\
$^3$Department of Plant Taxonomy and Ecology, E\"otv\"os University
H-1083 Budapest, Ludovika t\'er 2, Hungary}
\address{\em \today}

\address{
\centering{\medskip \em\begin{minipage}{15.4cm}
{}\qquad As a generalization of the 3-strategy Rock-Scissors-Paper game
dynamics in space, cyclical interaction models of six mutating species are 
studied on a square lattice, in which each species is supposed to have two
dominant, two subordinated and a neutral interacting partner. Depending on
their interaction topologies, these systems can be classified into four
(isomorphic) groups exhibiting significantly different behaviors as a
function of mutation rate. In three out of four cases three (or four)
species form defensive
alliances which maintain themselves in a self-organizing polydomain structure
via cyclic invasions. Varying the mutation rate this mechanism results in an
ordering phenomenon analogous to that of magnetic Ising systems.
\end{minipage}
}}
\maketitle\narrowtext

The Rock-Scissors-Paper (RSP) game is certainly one of the least sophisticated
games to play, yet one of the most interesting subjects of game theoretical
studies \cite{HS}. The rules of the game are indeed very simple: each of two
opponents choose a strategy from the strategy set {R, S, P}, neither knowing
the choice of the other. The strategies chosen compete; the winner is
determined according to a cyclic scheme of dominance: R beats S beats P
beats R. If the opponents choose the same strategy, the outcome is a draw.

The RSP game and its generalizations have received considerable attention
as models of cyclic interactions in biology and in economics \cite{MS,W}.
Added to the classical textbook example of the cyclic preference system
of mating partner choice in females of the lizard species {\it Uta stansburiana}
\cite{lizard},
a multi-species version of the RSP game has been used recently as a model
for cyclic interference competition among different strains of
bacteriocin-producing bacteria \cite{SCz,CzPH}.
The strategies are assumed to be genetically determined traits in most
biological applications: the individuals pass their strategy over to
their offspring (i.e., the strategies are heritable). The basic assumption
of such applications of the generalized RSP game as models of population
interactions is that pairs of individuals of a population repeatedly play
the RSP game (or a similar game of cyclical interaction topology) against
each other, receiving payoffs according to the outcome of the interaction.

The RSP game can be generalized in many different ways. Increasing the
number of strategies (species) by preserving the simple directed ring
structure for the graph of the topology of dominance relations is the
most straightforward generalization. Some theoretical aspects of cyclic
dominance in such models have already been thoroughly investigated
\cite{HS,BG,T94,FKB}. The analyses have justified that site invasions
based on payoff asymmetry (i.e., takovers of the loser's site by the
winner) maintain a self-organizing domain structure within a $d$-dimensional
lattice if $n$, the number of strategies (species) is less than a
($d$-dependent) threshold value for $d \geq 2$ \cite{FK}. It is emphasized
that these self-organizing structures develop only if system size exceeds
the typical size of single-species domains, otherwise one of the species
inevitably excludes all others from the lattice. Henceforth our analysis
will be restricted to sufficiently large systems in which drift due to
finite system size does not play a decisive role.

Further generalization of the multi-strategy RSP model is possible by
allowing for a reticulate topology of the still cyclic dominance graph,
so that each strategy can have more than one dominant and more than one
subordinated opponents within the strategy set. In a previous
paper \cite{SCz} we have studied a model of bacteriocin-mediated competition
among 9 strains of toxin-producing bacteria, which is an example to this
type of generalization. We also allowed for strains to mutate into each
other in that model, with mutation rates set according to loss rates of
toxin and resistance genes. Numerical analysis of the model has revealed
that this system undergoes a critical phase transition belonging to the
universality class of that of the three-state Potts model \cite{Potts,Wu},
if mutation rates decrease below a critical value. For high mutation rates,
all the nine species are present with the same probability. For low mutation
rates, however, one of three equivalent "coalitions" or " defensive alliances"
takes over in finite time, due to unlimited domain expansion within the
finite grid. The alliances consist of three species each, and their
stability is related to cyclic within-domain invasions \cite{T94}
providing protection against external intruders.

We have given the definition of a defensive alliance in \cite{SCz}, but
since it is the central concept of this paper as well, we repeat it here.
A defensive alliance is a directed circuit in the interaction graph, each
member of which is defended against its external dominant by its
within-alliance (internal) dominant. More precisely, if $a$, $b$ and
$c$ are the members of a defensive alliance ($a$ beats $b$ beats $c$
beats $a$), and $x$ is an external intruder dominant over $a$ ($x$
beats $a$), then $c$ is dominant over $x$ ($c$ beats $x$).

\begin{figure}\centerline{\epsfxsize=7.6cm
                          \epsfbox{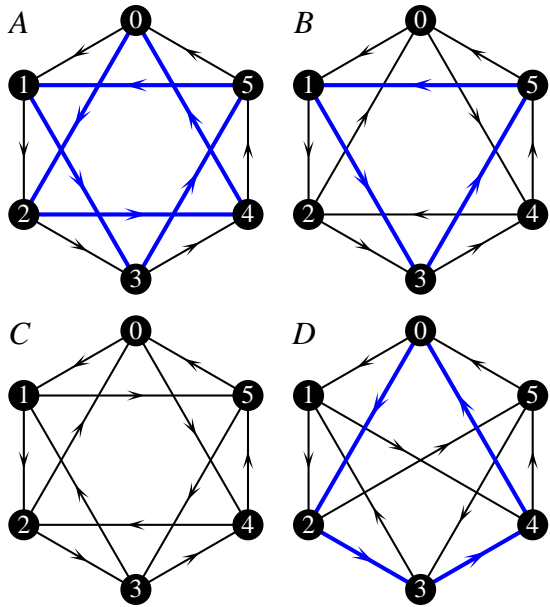}
                          \vspace*{0.5mm}   }
\caption{The interaction topologies are characterized by the directed
graphs $A$, $B$, $C$, and $D$. The bullets represent species labelled
by figures. The arrows show the direction of competitive invasion. The
species connected by thicker edges form defensive alliances.}
\label{fig:foodwebs}
\end{figure}

To specify somewhat more general criteria of stability in such systems
of cyclical competition in terms of the topology of the interaction graph,
the system needs to be further simplified. In this paper we reduce the number
of interacting strategies (species) to six, and require each species to have
exactly two dominant and two subordinated competitors among the remaining
five species. Any system satisfying these conditions can be characterized
by a directed graph and each one is isomorphic to one of the four graphs
shown in Fig.~\ref{fig:foodwebs}.

We have systematically tested these four classes of interaction topologies
for the emergence of defensive alliances in a way similar to that of the
previous, 9-species model. In the present spatial model each site of a square
lattice is occupied by one of the six species. System evolution is
governed by nearest neighbor invasion and local mutation. The latter means
that a site transforms itself to one of the two dominant species of the
actual resident species independently of neighborhood composition. For
simplicity the mutation rate $P$ is the same for all the
species, thus $P$ is the only parameter of the model.

The random sequential update consists of the following steps: 1) Random choice
of the focal site; 2) Decision on mutation: with probability $P$ a mutation
event occur on the focal site. Which of the two dominants replaces the focal
species is determined at random; 3) If no mutation occurs, one of the nearest
neighbors of the focal site is chosen
at random, and the species on the two sites play a round of the game. If one
dominates the other, both sites will be occupied by the dominant (i.e., the
dominant species invades the site of the other); nothing changes otherwise.

Notice that each graph of Fig.~\ref{fig:foodwebs} has a directed Hamiltonian
circuit positioned along the peripheries \cite{Deo}. Graphs $A$, $B$, and
$C$ have two additional three-edge circuits (with different directions)
composed of the remaining six "internal" edges, each of which embraces three
species potentially forming a defensive alliance.

In case $A$ the system has indeed just these two equivalent defensive
alliances consisting of species 0+2+4 and 1+3+5 respectively. Such an
association of the corresponding three species maintains a self-organizing
polydomain structure in the spatial model \cite{T94}, and cyclic invasions
prevent the invasion of species external to the alliance. For example, species
1 can invade the territory of species 2 in the 0+2+4 association, but species
0 is dominant over both 1 and 2. Consequently species 1, the external invador,
is soon abolished from the 0+2+4 domain by 0, the very same species which
dominates species 2 within the alliance. Mutatis mutandis, the same argument
applies to the other two (3 and 5) external invadors of alliance 0+2+4, and
the situation is completely symmetric with respect to the defensive mechanism
of the 1+3+5 alliance against 0, 2 and 4 as external invadors.

Graph $B$ admits only one such defensive alliance (1+3+5), and any
distinguished three-species association is missing from graph $C$. In this
latter case one can find eight eqivalent three-species circuits in the
corresponding directed graph, but neither is an alliance in the above sense.

The situation is quite different in graph $D$ which has four three-species
directed circuits (1+2+3, 3+4+5, 0+1+4 and 0+2+5), but neither of the
corresponding associations are protected against external invadors in
the sense mentioned above. However, this topology exhibits three directed
four-species circuits as well, and one of them (0+2+3+4) seems to be
self-protected.

It is worth mentioning here that graph $A$ involves three additional
four-species directed circuits which are also protected, i.e., it is not
straightforward to determine from the graph topologies alone which species
or associations would persist in the long run. We used Monte Carlo (MC)
analyses to predict the stationary distribution of competing species for
each of the four graphs. The MC simulations have been performed on a square
lattice of size $L \times L$ with periodic boundary conditions for different
mutation rates ($P$) and interaction topologies. System size varied from
$L=300$ to 2000. We used larger grids to suppress the undesired consequences
of increased fluctuations we have found in some cases. The initial
configurations have been assembled at random, each species having the same
chance to occupy a site. After some thermalization we have recorded the
concentrations of the species, averaging over a sampling time ranging
from $10^4$ to $5 \cdot 10^5$ MC steps per site, as required by the actual
fluctuations.

\begin{figure}
\centerline{\epsfxsize=7.5cm
                          \epsfbox{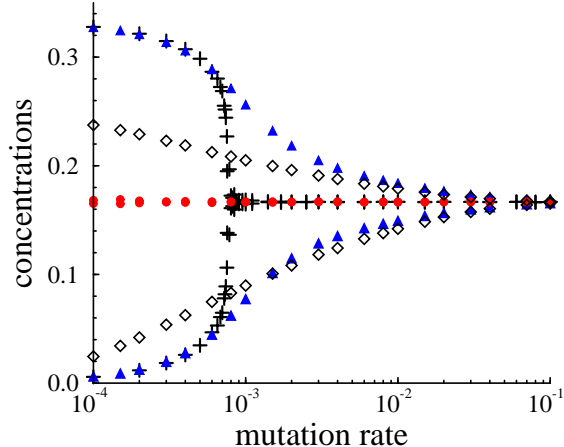}
                          \vspace*{0.5mm}   }
\caption{Average stationary concentrations of species as a function of mutation
rate for graphs $A$ (pluses), $B$ (closed triangles), $C$ (bullets), and $D$
(open diamonds).}
\label{fig:cmr}
\end{figure}

The results of the simulations are summarized in Fig.~\ref{fig:cmr}.
The simplest case is $C$, where all the species are present in equal
abundances in the stationary state if $P>0$ and the lattice is sufficiently
large. System $A$ behaves similarly if the mutation rate exceeds a critical
value ($P > P_c$). Below this transition point one of the defensive alliances
(0+2+4 or 1+3+5) takes over. In each of graphs $B$ and $D$ there exists only
one defensive alliance whose dominance increases smoothly as the mutation
rate decreases towards 0.

Figure~\ref{fig:cmr} clearly demonstrates that in the limit $P \to 1$
(when mutation governs system evolution alone) the species concentrations
become equal. Apart from this limiting case, the species belonging to the
dominating alliance (if there exists such) are present in the same abundance
(concentration). That is, in the limit $P \to 0$ the "allied" species
concentrations tend to either 1/3 (for graphs $A$ and $B$) or 1/4 ($D$).

For model $A$ the detailed analysis shows that the $P$-dependence of average
concentrations can be described by introducing an order parameter as
\begin{eqnarray}
\langle c_0 \rangle &=& \langle c_2 \rangle =\langle c_4 \rangle =
{1 \over 6} (1 \pm m) \, , \nonumber \\
\langle c_1 \rangle &=& \langle c_3 \rangle =\langle c_5 \rangle
={1 \over 6} (1 \mp m) \,
\label{eq:op}
\end{eqnarray}
where $m=0$ if $P>P_c$, otherwise the order parameter is larger than zero
($0 < m \leq 1$). The results of our MC simulations support that the
order parameter follows a power law behavior ($m \propto (P_c-P)^{\beta}$)
below the critical point in the close vicinity of $P_c$. The best fit is
found for $\beta=0.127(8)$ and $P_c=0.0007515(5)$.
Within the statistical error, this value of $\beta$
agrees with that found for the two-dimensional Ising model
($\beta_{\rm Ising}=1/8$) \cite{Onsager,Wu}. In the present case the ordered
phase is twofold degenerated because only one of the two defensive alliances
survives (with the same probabiliy) in the limit $P \to 0$. It is therefore
not surprising that the corresponding critical transition belongs to the
universality class of the Ising model \cite{Grins}.

In order to have further numerical evidences concerning this critical
transition we have studied the concentration fluctuation defined as
\begin{equation}
\chi = L^2 \sum_s \langle (c_s - \langle c_s \rangle)^2 \rangle
\end{equation}
where summation runs over the species ($s=0, \ldots ,5$). For model $A$
this quantity diverges at the critical transition on large lattices.
More precisely, our MC data can be well approximated by a power law;
$\chi \approx |P -P_c|^{-\gamma}$ in the vicinity of $P_c$. Below and
above the critical point, numerical fitting yields $\gamma=1.0(2)$ and
$\gamma^{\prime}=1.1(2)$, respectively, in agreement with the theoretical
expectation $\gamma_{\rm Ising}=\gamma^{\prime}_{\rm Ising}=1$ \cite{Wu}.

\begin{figure}
\centerline{\epsfxsize=7.5cm
                          \epsfbox{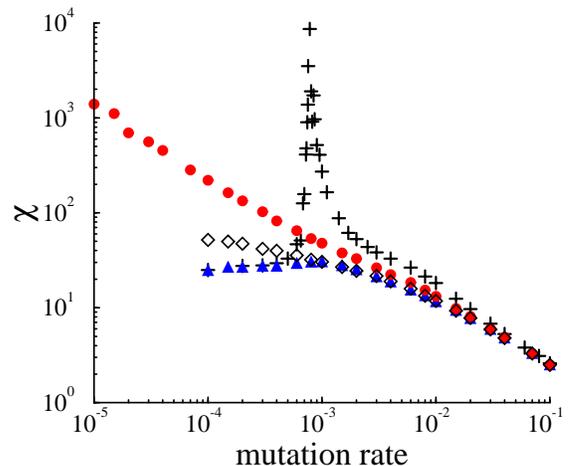}
                          \vspace*{0.5mm}   }
\caption{Concentration fluctuations vs. mutation rate for the four models.
The MC data are indicated by the same symbols as in Fig.~\ref{fig:cmr}.}
\label{fig:flmr}
\end{figure}

Figure~\ref{fig:flmr} compares the $P$-dependence of $\chi$ we obtained
by MC simulations for the different interaction graphs. $\chi$ remains
low for graphs $B$ and $D$, but this quantity diverges for $C$, in which
case the power law fit yields $\chi \propto P^{- \gamma}$ with
$\gamma = 0.72(3)$ in the limit $P \to 0$. This behavior implies a
critical transition at $P=0$. Consequently, in the absence of mutation
some ordering process is expected which is studied via considering the
time-dependence of correlation length at zero mutation rate \cite{Bray}.

If the system is started from a random initial state, then the visualization
of the species distribution exhibits an ordering process for all these models
at $P=0$. The situation is easy to understand for the graphs $B$ and $D$.
In these cases the models admit only a single defensive alliance that emerges
in the whole system homogeneously. Therefore the species concentrations tend
exponentially towards the stationary values plotted in Fig.~\ref{fig:cmr}.
For graph $A$, however, there exist two equivalent defensive alliances both
forming domains of increasing size, in close analogy to the Ising models for
zero magnetic field. It is emphasized that inside the domain of a defensive
alliance the cyclic invasions maintain a self-organizing three-color pattern,
just as in another multicycle RSP model \cite{SCz}.
The numerical analysis of the "first zero" of the equal-time pair correlation
function [$C(\xi_0(t),t)=0$] supports that the deduced characteristic linear
size of domains follows the usual growth law, namely,
$\xi_0(t) \propto \sqrt{t}$ \cite{Bray}.

As mentioned above, graph $C$ has eight equivalent three-species cycles and
each one can dominate the system with the same chance in the stationary state
for finite lattice size if $P=0$. The visualization of species distribution
again indicates spatial ordering. The most striking feature is the extinction
of several species from certain regions, and that the size of such regions
increases with time. The quantitative analysis of the equal-time pair
correlation function confirms the expected results, i.e., the corresponding
correlation length increases as $\xi (t) \propto \sqrt{t}$. Here the
correlation length is derived by fitting an exponential function
($e^{x/\xi (t)}$) to the MC data of $C(x,t)$ obtained by averaging over
five runs for $L=5000$.

In all these finite-size systems the number of surviving species is reduced
to 3 (or 4 for graph $D$) via a domain growth mechanism, and the coexistence
of the surviving species is maintained by cyclic invasions. Although the
interaction graphs studied may admit many possible cycles, we found that
evolution favors certain topological structures that we call "defensive
alliances" capable of protecting themselves from external invasion. System
behavior can be predicted by thoroughly studying the potential alliances
within the interaction graph of the model in question. In two of four graphs
($B$ and $D$) there is only a single defensive alliance, and that one
inevitably dominates the system in spite of the permanent attacks by mutants
and external invadors. If the interaction graph admits two equivalent defensive
alliances ($A$), the population undergoes a critical phase transition
(accompanied with spontaneous symmetry breaking) when varying the mutation
rate. This means that above a critical rate of mutations ($P > P_c$) all the
species persist with the same average concentration in the stationary state,
whereas below $P_c$ the members of one of the defensive alliances take over,
driving all other species extinct. It is interesting that the 4-species
alliances of graph $A$ do not show up at all in the system. The reason for
this might be twofold: 1) the 4-species alliances are overlapping, that is,
each species is a member of more than one alliance; 2) it is topologically
easier to maintain a 3-species alliance within a 2D lattice, because the
neighborhood relations of the allies chasing each other is more probable
to persist in space among 3 than among 4 species. In the fourth case ($C$)
the graph has eight equivalent three-species cycles whose self-organizing
patterns are easily destroyed by external invadors and mutants, because none
of them form defensive alliances. Consequently, external invasions and the
appearance of mutants stop domain growth (which can be observed locally for
$P=0$) and maintain a global state with equal species concentrations. 
Just as in many other models of ecological processes \cite{CzT}, pattern 
formation plays a crucial role in the self-protection of defensive alliances: 
an external intruder is quickly abolished from the domain of an
alliance, because the internal enemy of the intruder is close by: it is
chasing the attacked species. Similar spatio-temporal patterns have been
observed in other models (e.g., in models of forest fires \cite{ffm},
evolutionary games \cite{epdg3s}, etc.) in which the invasion speeds were
different for different pairs of interacting species. This fact implies
that some of our predictions regarding these simple models can be generalized
to systems with different invasion and mutation rates as well.

\acknowledgements
Support from the Hungarian National Research Fund (T-33098 and T-25793)
is acknowledged.


\begin{references}

\bibitem{HS}J. Hofbauer and K. Sigmund, {\it Evolutionary Games and Population
Dynamics} (Cambridge University Press, Cambridge, 1998).

\bibitem{MS}J. Maynard Smith, {\it Evolution and Theory of Games}
(Cambridge University Press, Cambridge, 1982).

\bibitem{W}J. W. Weibull, {\it Evolutionary Game Theory} 
(Oxford University Press, Oxford, 1993).

\bibitem{lizard}B. Sinervo and C. M. Lively, Nature {\bf 380}, 240 (1996);
J. Maynard Smith, ibid. 198.

\bibitem{SCz}G. Szab\'o and T. Cz\'ar\'an, cond-mat/0008311.

\bibitem{CzPH}T. Cz\'ar\'an, R. F. Hoekstra, and L. Pagie (to be published).

\bibitem{BG}M. Bramson and D. Griffeath, Ann.\ Prob.\ {\bf 17}, 26 (1989).

\bibitem{T94}K. Tainaka, Phys.\ Rev.\ E {\bf 50}, 3401 (1994).

\bibitem{FKB}L. Frachebourg, P. L. Krapivsky, and E. Ben-Naim, 
Phys.\ Rev.\ E {\bf 54}, 6186 (1996).

\bibitem{FK}L. Frachebourg and P. L. Krapivsky,
J. Phys.\ A: Math.\ Gen.\ {\bf 31}, L287 (1998).

\bibitem{Potts}R. B. Potts, Proc.\ Camb.\ Phil.\ Soc.\ {\bf 49}, 106 (1952).

\bibitem{Wu}F. Y. Wu, Rev.\ Mod.\ Phys.\ {\bf 54}, 235 (1982).

\bibitem{Deo}Narsingh Deo, {\it Graph Theory with Applications to
Engineering and Computer Science} (Prentice-Hall, Englewood Cliffs, 1974).

\bibitem{Onsager}L. Onsager, Phys.\ Rev.\ {\bf 65}, 117 (1944).

\bibitem{Grins}G. Grinstein, C. Jayaprakash, and Yu He, Phys.\ Rev.\ Lett.\ 
{\bf 55}, 2527 (1985).

\bibitem{Bray}A. J. Bray, Adv.\ Phys.\ {\bf 43}, 357 (1994).

\bibitem{CzT}T. Cz\'ar\'an, {\it Spatiotemporal Models of Population and
Community Dynamics} (Chapman and Hall, London, 1998).

\bibitem{ffm}P. Grassberger and H. Kantz, J.\ Stat.\ Phys.\ {\bf 63}, 685 
(1991); B. Drossel and F. Schwabl, Phys.\ Rev.\ Lett.\ {\bf 69},1629 (1992);
J. E. S. Socolar, G. Grinstein, and C. Jayaprakash, Phys.\ Rev.\ E {\bf 47}, 
2366 (1993).

\bibitem{epdg3s}G. Szab\'o, T. Antal, P. Szab\'o, and M. Droz,
Phys.\ Rev.\ E {\bf 62}, 1095 (2000).

\end{references}
\end{document}